# Superconductivity at 4K in Pd deficient layered $Ta_2Pd_xS_6$

Brajesh Tiwari[1], Babu Baijnath Prasad[1,2], Rajveer Jha[1], Dharmendra Kumar Singh[1] and V.P.S. Awana[1,*]

[1]*National Physical Laboratory, New Delhi-110012, India*
[2]*Sri Murli Manohar Town Post Graduate College Ballia. 277001, India*

Strong spin-orbit coupled 4d and 5d transition metal based compounds such as $(Nb/Ta)_2Pd_x(S/Se/Te)_y$ have gained renewed research interests due to their exotic superconducting properties, such as exceptionally large upper critical fields, being surpassing the Pauli paramagnetic limit, $H_p \sim 1.84 T_c$, their multi band nature and pressure independent superconducting transition temperature ($T_c$) [1-8]. In a very recent experimental observations Lu et. al [6] first reported that $Ta_2Pd_xS_5$ exhibit metal to superconducting transition at around 6K with three times higher upper critical field, $H_{c2}(0)$, to that as constrained by Pauli paramagnetic limit. This is theoretically explained by considering the multiband superconductivity in strong coupling regime [7].

The $Ta_2Pd_xS_6$ compounds were first synthesized and characterized in mid 1980's [9,10]. Following the available reports [1-8] on superconductivity of (Nb/Ta)Pd(S/Se/Te) compounds, the identification of right phase and composition of them seems to be the major issue yet. Further, the role of Pd-centered square planer structure needs to be probed in detail to comprehend the extraordinary physical properties in Pd-based chalcogenide superconductor [1-8].

Here in this short note, we report on the low dimensional 4d and 5d transition metals-chalcogenide based compounds i.e., $Ta_2Pd_xS_6$, showing semiconducting to superconducting transition around 4K with upper critical field outside Pauli paramagnetic limit. On the other hand in $Ta_2Pd_xS_5$ phase, a metallic normal state with $T_c \sim$ 6K is seen [6]. It seems couple of different superconducting phases do exist in these new set of compounds. Our short note in this regards is thought provoking, asking to explore various unearthed possible superconducting phases in $(Nb/Ta)_2Pd_x(S/Se/Te)_y$ systems.

In this study, we focus on the polycrystalline samples with nominal composition $Ta_2Pd_xS_6$; x= 0.97, 0.94. The stoichiometric ratio of elemental powder of Ta, Pd and S in 2:x:6 were mixed thoroughly in Ar-controlled glove-box and pelletized by uniaxial 100 kg/cm$^2$ stress. The pellets were sealed separately in evacuated quartz tubes and heated to 800$^o$C and 850$^o$C for 24h with intermediate grinding. The obtained samples are used for structural and transport measurements. The structure of the samples were identified by Rietveld refinement of powder XRD pattern recorded using Rigaku made machine with Cu Kα (λ=1.54Å) x-ray source. Resistivity of bulk samples were measured by standard linear four probe method on Quantum Design make Physical Property Measurement System (PPMS).

Figure 1(a) shows the XRD patterns for $Ta_2Pd_{0.97}S_6$ and $Ta_2Pd_{0.94}S_6$ and their corresponding Rietveld refinements with two of their monoclinic phases with space group C2/m having two different stoichiometries i.e., $Ta_2Pd_xS_6$ and $Ta_2Pd_xS_5$ (x=0.97 and 0.94).

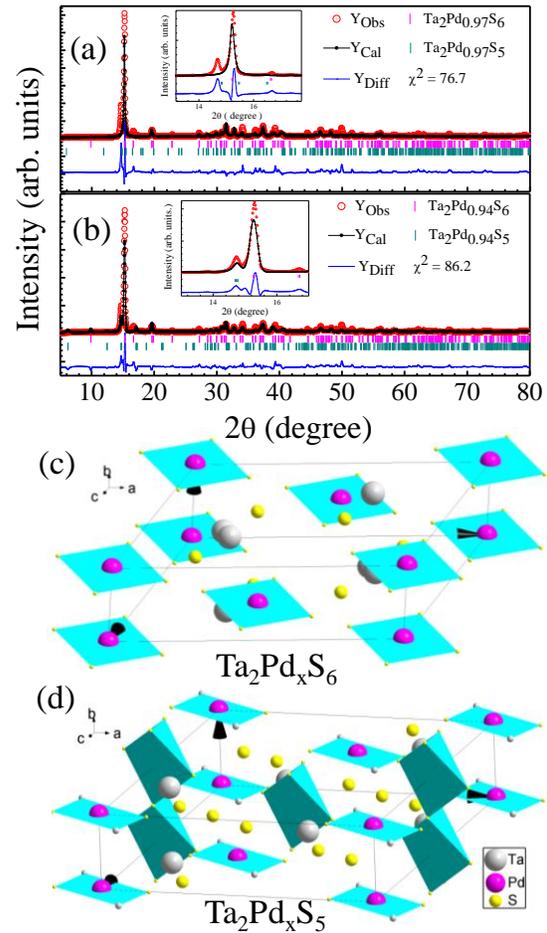

Figure 1 (Color online). Rietveld refinement for powder XRD patterns of compounds (a) $Ta_2Pd_{0.97}S_6$ and (b) $Ta_2Pd_{0.94}S_6$ with monoclinic space group C2/m. The XRD patterns are refined simultaneously for the two different crystal structures (c) $Ta_2Pd_xS_6$ (magenta bars) and (d) $Ta_2Pd_xS_5$ (cyan bars), respectively.

It is observed that though the samples were not single phase, the major phase (> 80% by volume fraction) in these samples belongs to composition

$Ta_2Pd_xS_6$, along with minor $Ta_2Pd_xS_5$ phase. The corresponding crystal structures are shown in figure 1(b) and (c) respectively. The obtained lattice parameters for $Ta_2Pd_{0.94}S_6$ phase are $a=11.79$Å, $b=3.27$Å, $c=9.93$Å, $\beta=115.6°$ while for $Ta_2Pd_{0.94}S_5$ the same are $a=12.31$Å, $b=3.29$Å, $c=14.56$Å, $\beta=102.7°$. A small increase in $a$ and $c$ lattice parameters is found for $Ta_2Pd_{0.97}S_6$, which are close to earlier reported values for the same phase [10].

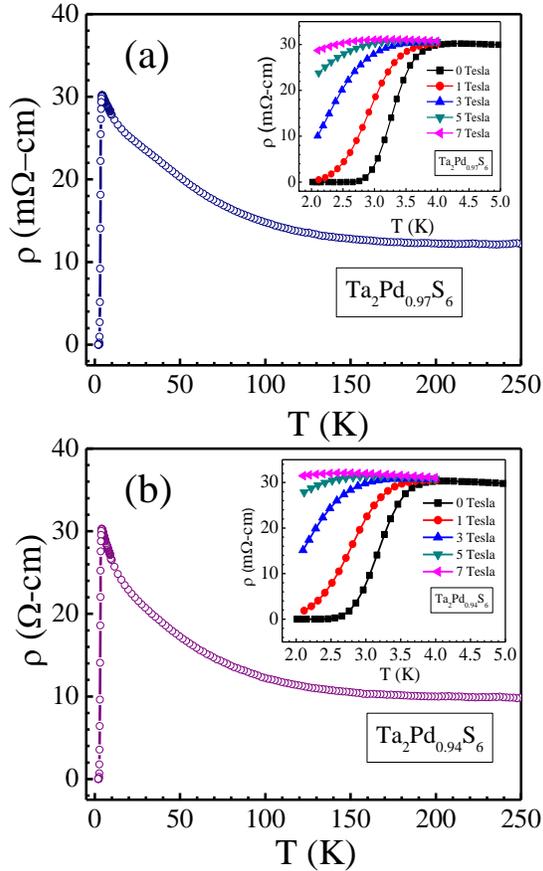

**Figure 2** (Color online). Temperature (T) dependences of resistivity (ρ) for compounds (a) $Ta_2Pd_{0.97}S_6$ and (b) $Ta_2Pd_{0.94}S_6$, whereas insets show ρ verses T at different applied magnetic fields upto 7 Tesla, respectively.

Figure 2 shows the temperature dependences of electrical resistivity for compounds (a) $Ta_2Pd_{0.97}S_6$ and (b) $Ta_2Pd_{0.94}S_6$ down to 2 K. A clear semiconducting normal state can be observed in variance to recent report by Lu et.al [6] which observed metallic normal state in $Ta_2Pd_xS_5$ stoichiometry. $Ta_2Pd_{0.97}S_6$ (and $Ta_2Pd_{0.94}S_6$) shows superconducting transition with onset temperature $T_c^{onset} = 3.96K$ (3.76K) and estimated $dH_{c2}/dT$ based on 90% criteria of normal state resistivity is -4.43Tesla/K (and -3.53Tesla/K). The estimated $H_{c2}(0)$ is 12Tesla (and 9.2Tesla), which is clearly above Pauli's paramagnetic limit of 7.3Tesla for these 4K superconductors. For the determination of upper critical field we used the formula $H_{c2}(0) = -0.69 T_c \left(\frac{dH_{c2}}{dT}\right)_{T=T_c}$ based on model developed by Werthamer Helfand and Hohenberg (WHH) true for BCS type superconductors [11]. These observations suggest entirely different and a new superconducting phase in Pd-based compound $Ta_2Pd_xS_6$. The detailed study is underway on role of sulfur stoichiometry on the structure and physical properties of $Ta_2Pd_xS_{6-\delta}$ based superconductors.

In summary, we have synthesized Pd-deficient monoclinic compounds with major $Ta_2Pd_xS_6$ stoichiometry. We have shown by temperature dependent resistivity measurements that $Ta_2Pd_xS_6$ have normal state semiconducting to superconducting transition at about 4 K. It is observed that these compounds are rather robust against external magnetic field. Observation of superconductivity in heavy transition metal-based chalcogenide may add to new dimensions in superconducting research.

**Acknowledgement:** Authors would like to thank their Director NPL India and for financial support by *DAE-SRC* scheme on search for new superconductors. BBP thanks IASs, India for summer research fellowship.

**References:**
[1]. Q. Zhang, G. Li, D. Rhodes, A. Kiswandhi, T. Besara, B. Zeng, J. Sun, T. Siegrist, M. D. Johannes, and L. Balicas, *Sci. Rep.,* **3**, 1446 (2013).
[2]. R. Jha, B. Tiwari, P. Rani, H. Kishan and V. P. S. Awana, *J. Appl. Phys.* **115**, 213903 (2014).
[3]. C. Q. Niu, J. H. Yang, Y. K. Li, B. Chen, N. Zhou, J. Chen, L. L. Jiang, B. Chen, X. X. Yang, C. Cao, J. Dai, and X. Xu *Phys. Rev. B*, **88**, 104507 (2013)
[4]. H. Yu, M. Zuo, L. Zhang, S. Tan, C. Zhang and Y. Zhang, *J. Am. Chem. Soc.*, **135** (35), 12987 (2013).
[5]. S. Khim, B. Lee, Ki-Y. Choi, B.-Gu Jeon, D. H. Jang, D. Patil, S. Patil, R. Kim, E. S. Choi, S. Lee, J. Yu and K. H. Kim, *New J. of Phys.,* **15**, 123031 (2013).
[6]. Y. Lu, T. Takayama, A. F. Bangura, Y. Katsura, D. Hashizume and H. Takagi, *J. Phys. Soc. Jpn.,* **83,** 023702 (2014).
[7]. D. J. Singh, *Phys. Rev. B*, **88**, 174508 (2013).
[8]. W. H. Jiao, Z. T. Tang, Y. L. Sun, Y. Liu, Q. Tao, C. M. Feng, Y. W. Zeng, Z. A. Xu and G. H. Cao *J. Am. Chem. Soc.*, **136** (4), 1284 (2014).
[9]. P. J. Squattritto, S. A. Sunshine and J. A. Ibers, J. Solid State Chem. **64**, 261 (1986).
[10]. D. A. Keszler, P. J. Squattrito, N. E. Brese, J. A. Ibers, S. Maoyu, and L. Jiaxi, *Inorg. Chem.,* **24**, 3063 (1985).
[11]. N. R. Werthamer, E. Helfand, and P. C. Hohenberg, *Phys. Rev.*, **147**, 295 (1966).